\documentclass{JHEP3}
\usepackage{textcomp}
\usepackage{amsmath, amssymb}

\title{Covariant propagator in AdS$_5\times$S$^5$ superspace}

\author{
Peng Dai \footnote{Email: peng.dai@stonybrook.edu}$^{~1}$,
Ru-Nan Huang \footnote{Email: huangrn04@lzu.cn}$^{~2,1}$, and
Warren Siegel \footnote{Email: siegel@insti.physics.sunysb.edu}$^{~1}$
\\ \\ \\
\it $^1$ C.N. Yang Institute for Theoretical Physics,\\
\it Stony Brook University,\\
\it Stony Brook, NY 11794-3840, USA
\\ \\
\it $^2$ Institute of Theoretical Physics,\\
\it Lanzhou University,\\
\it Lanzhou 730000, People's Republic of China}

\abstract{
We give an explicit superspace propagator for the chiral scalar field strength of 10D IIB supergravity on an AdS$_5\times$S$^5$ background. Because this space is conformally flat, the propagator is very simple, almost identical to that of flat space. We also give an explicit expansion over the Kaluza-Klein modes of S$^5$. The fact that the full propagator is so much simpler suggests that, as in 2D conformal field theory, AdS/CFT calculations would be simpler without a mode expansion.}


\preprint{YITP-SB-09-37}

\begin{document}

\section{Introduction}
Two of the most interesting results from the anti-de Sitter/conformal field theory correspondence are implied by string theory S-matrices with external states on either the boundary \cite{boundary} or the horizon \cite{horizon} of AdS$_{5}$. Although supergraph methods are known to simplify calculations and extend results in ``lower-dimensional'' supersymmetric theories, they have not yet been applied to this 10D case. In particular, in the boundary case the emphasis has been placed on the string ground states, IIB supergravity, which describe all the BPS states of the CFT. This suggests that IIB supergravity supergraphs would be a useful approach to extend results in this area, and facilitate comparison to the corresponding 4D N=4 Yang-Mills states, also described in superspace. For the latter purpose, it is useful to not expand in the Kaluza-Klein modes of the sphere, since some of those coordinates correspond to R-symmetry coordinates of the 4D theory.

We find a further reason for avoiding this expansion is that it introduces unnecessary complication into the Feynman-Witten graph rules. In particular, we show here that one of the relevant propagators, that for the chiral scalar superfield strength \cite{chiral} (dilaton +$i$ axion at $\theta=0$), has an almost obvious form before expansion. After expansion, the usual hypergeometric functions are obtained. (We also give the explicit form of the latter in terms of elementary functions, which is possible because their integer and half-integer conformal weights are protected.)

The outline of the paper is as follows: In the next section we look at the purely bosonic case, and derive the scalar propagator on AdS${}_n\times$S${}^n$ from flat space by Weyl scale transformation. This is possible because (1) the space is conformally flat (the Weyl tensor vanishes), and (2) the scalar curvature also vanishes, so we can ignore the fact that the scalar of interest has no improvement term (and hence isn't truly conformal in curved space). We next give the explicit Kaluza-Klein expansion of the propagator. In section 4 we generalize to supersymmetry (in 10-dimensional superspace) by constructing the appropriate coset space \cite{coset}, using the coset methods of \cite{Roiban}, and defining the action of the covariant derivatives. In the next section we define the supersymmetrized invariant lengths $s_x$ of AdS${}_5$ and $s_y$ of S${}^5$ by purely algebraic methods, i.e., without a choice of (super)coordinate system. Applying these same methods in section 6,  we solve the Klein-Gordon equation to find the propagator in superspace: $$ \Delta = {3!\over (s_x - s_y)^4} $$ which has the same form as the above bosonic result.  In the following section we solve the $\kappa$-symmetry constraint, which applies no additional conditions (because the chirality constraint has already been solved), but is a useful consistency check.  A simple coordinate choice is given in section 8. We conclude with proposals for interactions.

%%%%%%%%%%%%%%%%%%%%%%%%%%%%%%%%%%%%%%%%%%%%%
\section{The propagator in AdS$_n\times$S$^n$}
We shall now calculate the propagator of the massless scalar field in AdS$_n\times$S$^n$. Using Poincar\'e coordinates for the Euclidean AdS$_n$, the line element of AdS$_n\times$S$^n$ can be written as
\begin{subequations}
\begin{align}
ds=&\frac{dx^2+dx_0^2}{x_0^2}+d\Omega_n\\
=&\frac{dx^2+dx_0^2}{x_0^2}+d\hat{Y}^2\nonumber\\
=&\frac{dx^2+dY^2}{x_0^2}\label{conformal-flat line element}
\end{align}
\end{subequations}
where $\hat{Y}$ are unit vectors in $\mathbb{R}^{n+1}$ and $Y\equiv x_0\hat{Y}$. From eq.\eqref{conformal-flat line element}, it is clear that a Weyl transformation can give us the flat metric
\begin{equation}
g_{\mu\nu}\rightarrow x_0^2g_{\mu\nu}=\eta_{\mu\nu}\label{Weyl transformation of the metric}
\end{equation}

The action of the massless scalar field in AdS$_n\times$S$^n$ is
\begin{equation}
S=\int d^{2n}z\sqrt{g}\phi\square_g\phi
\end{equation}
Performing the Weyl transformation \eqref{Weyl transformation of the metric} and the following field scaling
\begin{equation}
\phi\rightarrow x_0^{-n+1}\phi=\phi^\prime
\end{equation}
we can bring the action into flat space
\begin{equation}
S=\int d^{2n}z\phi^\prime\square_\eta\phi^\prime
\end{equation}

It is easy to get the flat-space propagator
\begin{equation}
\Delta_\eta(1,2)
\equiv \langle\phi^\prime(1)\phi^\prime(2)\rangle
=(n-2)!\frac{1}{\left[\frac{1}{2}(x_1-x_2)^2+\frac{1}{2}(Y_1-Y_2)^2\right]^{n-1}}
\end{equation}
which is normalized such that
\begin{equation}
-\frac{1}{2}\square_\eta\Delta_\eta(1,2)=(2\pi)^n\delta(1,2)
\end{equation}
Weyl scaling back, we get the propagator in AdS$_n\times$S$^n$ as
\begin{subequations}
\begin{align}
\Delta_g(1,2)\equiv&\langle\phi(1)\phi(2)\rangle
=(n-2)!\frac{x_{01}^{n-1}x_{02}^{n-1}}{\left[\frac{1}{2}(x_1-x_2)^2+\frac{1}{2}(Y_1-Y_2)^2\right]^{n-1}}\\
=&(n-2)!\left[\frac{(x_1-x_2)^2}{2x_{01}x_{02}}
+\frac{(x_{01}-x_{02})^2+2x_{01}x_{02}-2x_{01}x_{02}\hat{Y}_1\cdot\hat{Y}_2}{2x_{01}x_{02}}\right]^{-n+1}\nonumber\\
=&(n-2)!\left[1+\frac{(x_1-x_2)^2+(x_{01}-x_{02})^2}{2x_{01}x_{02}}-\hat{Y}_1\cdot\hat{Y}_2\right]^{-n+1}
\end{align}
\end{subequations}

Defining the $(n+1)$ dimensional unit vector
\begin{equation}
\hat{X}=\frac{1}{x_0}\left(1,\tfrac{1}{2}(x^2+x_0^2),x_i\right)
\end{equation}
and the corresponding metric
\begin{equation}
\eta^\prime=\left(\begin{array}{ccccc}
&-1\\ -1\\ && +1\\ &&&\ddots\\ &&&&+1
\end{array}\right)
\end{equation}
then the hyperbolic distance $d_{12}$ defined on AdS$_n$ is given by
\begin{equation}
s_x(1,2)\equiv-\hat{X}_1\cdot\hat{X}_2
=1+\frac{(x_1-x_2)^2+(x_{01}-x_{02})^2}{2x_{01}x_{02}}
=\cosh d_{12}
\end{equation}

We also have the spherical distance $\theta_{12}$ on S$^n$
\begin{equation}
s_y(1,2)\equiv\hat{Y}_1\cdot\hat{Y}_2=\cos\theta_{12}
\end{equation}
thus the massless scalar propagator in AdS$_n\times$S$^n$ can be written as
\begin{equation}
\Delta_{2n}(s_x,s_y)=\frac{(n-2)!}{(s_x-s_y)^{n-1}}\label{propagator in AdS_n x S^n}
\end{equation}

%%%%%%%%%%%%%%%%%%%%%%%%%%%%%%%%%%%%%%%%%%%%%
\section{Expansion over Kaluza-Klein modes}
We shall now derive the expansion of the propagator \eqref{propagator in AdS_n x S^n} over the KK modes of S$^n$.

The Laplacians on AdS$_n$ and S$^n$ can be respectively given by
\begin{equation}
\square_{AdS_n}=x_0^2\partial_0^2+x_0^2\sum_{i=1}^{n-1}\partial_i^2-(n-2)x_0\partial_0
\end{equation}
and
\begin{equation}
\hat{\square}_{S^n}=\frac{1}{\sin^{n-1}\theta}\frac{\partial}{\partial\theta}\left(\sin^{n-1}\theta\frac{\partial}{\partial\theta}\right)
+\frac{1}{\sin^2\theta}\hat{\square}_{S^{n-1}}
\end{equation}
Noting that $\Delta_{2n}$ is a function of only $s_x$ and $s_y$, we get that
\begin{equation}
\square_{AdS_n}\Delta_{2n}=\left[(s_x^2-1)\frac{\partial^2}{\partial s_x^2}+ns_x\frac{\partial}{\partial s_x}\right]\Delta_{2n}
\end{equation}
and
\begin{equation}
\hat{\square}_{S^n}\Delta_{2n}=-\left[(s_y^2-1)\frac{\partial^2}{\partial s_y^2}+ns_y\frac{\partial}{\partial s_y}\right]\Delta_{2n}
\end{equation}
Then the KG equation
\begin{equation}
\square\Delta_{2n}=\left(\square_{AdS_n}+\hat{\square}_{S^n}\right)\Delta_{2n}
\end{equation}
can be written as
\begin{equation}
\left[(s_x^2-1)\frac{\partial^2}{\partial s_x^2}+ns_x\frac{\partial}{\partial s_x}
-(s_y^2-1)\frac{\partial^2}{\partial s_y^2}-ns_y\frac{\partial}{\partial s_y}\right]\Delta_{2n}=0\label{KG equation in AdS_n x S^n (2)}
\end{equation}
When $s_x,s_y\neq1$, i.e., point 1 and 2 do not coincide, one can check that the propagator \eqref{propagator in AdS_n x S^n} indeed satisfies eq.\eqref{KG equation in AdS_n x S^n (2)}. Now, under such an assumption, let us solve eq.\eqref{KG equation in AdS_n x S^n (2)} directly.

Eq.\eqref{KG equation in AdS_n x S^n (2)} gives us the following identical differential equations with respect to $s_y$ and $s_x$, respectively
\begin{equation}
\left[(1-s_y^2)\frac{\partial^2}{\partial s_y^2}-ns_y\frac{\partial}{\partial s_y}+l(l+n-1)\right]\Delta_{2n}=0\label{Gegenbauer eq. with s_y}
\end{equation}
\begin{equation}
\left[(1-s_x^2)\frac{\partial^2}{\partial s_x^2}-ns_x\frac{\partial}{\partial s_x}+l(l+n-1)\right]\Delta_{2n}=0\label{Gegenbauer eq. with s_x}
\end{equation}
where $l=0,1,2, \dotsc$, and eq.\eqref{Gegenbauer eq. with s_y} is the eigenequation for spherical harmonics. These equations are known as the Gegenbauer differential equations.

For eq.\eqref{Gegenbauer eq. with s_y}, we know that $s_y\in[-1,1]$ and its solution is the Gegenbauer polynomial
\begin{equation}
C_l^{(\frac{n-1}{2})}(s_y)
=\binom{l+n-2}{n-2}F\left(-l,l+n-1,\frac{n}{2},\frac{1-s_y}{2}\right)
\end{equation}
where $F(a,b,c,z)$ is the hypergeometric function and $C_l^{(\frac{n-1}{2})}(s_y)$ is normalized such that $C_0^{(\frac{n-1}{2})}(s_y)=1$.

For eq.\eqref{Gegenbauer eq. with s_x}, we know that $s_x\in[1,\infty)$, and the boundary condition
\begin{equation}
s_x\rightarrow1, \quad \Delta_{2n}\rightarrow\infty; \quad s_x\rightarrow\infty, \quad \Delta_{2n}\rightarrow0
\end{equation}
tells us its solution is the Gegenbauer function of the second kind $D_l^{(\frac{n-1}{2})}(s_x)$:

1) If $n$ is odd
\begin{multline}
D_l^{(\frac{n-1}{2})}(s_x)
=\frac{2^{\frac{n}{2}}}{\sqrt\pi}\Gamma\left(\frac{n-1}{2}\right)
\left[(-)^{\frac{n-1}{2}}\frac{\pi}{2}C_l^{(\frac{n-1}{2})}(s_x)\right.\\
\left.+2^{n-4}\frac{\left[\Gamma(\frac{n}{2}-1)\right]^2}{\Gamma(n-2)}(s_x^2-1)^{-\frac{n}{2}+1}
F\left(-l-n+2,l+1,-\frac{n}{2}+2,\frac{1-s_x}{2}\right)\right]
\end{multline}

2) If $n$ is even
\begin{multline}
D_l^{(\frac{n-1}{2})}(s_x)
=\frac{\sqrt\pi}{2^{l+\frac{n}{2}-1}}\frac{\Gamma(l+n-1)}{\Gamma(l+\frac{n+1}{2})}\\
\times(s_x-1)^{-l-\frac{n}{2}}(s_x+1)^{-\frac{n}{2}+1}F\left(l+1,l+\frac{n}{2},2l+n,\left(\frac{1-s_x}{2}\right)^{-1}\right)
\end{multline}
Using the following formula (\cite{Erd}, p.69)
\begin{multline}
F\left(l+1,l+\frac{n}{2},2l+n,z\right)=\\
\frac{(-)^{\frac{n}{2}-1}(2l+n-1)!}{l![(l+\frac{n}{2}-1)!]^2(l+n-2)!}\frac{d^{l+\frac{n}{2}-1}}{dz^{l+\frac{n}{2}-1}}
\left[(1-z)^{l+n-2}\frac{d^{l+\frac{n}{2}-1}}{dz^{l+\frac{n}{2}-1}}\left(\frac{1}{z}\ln\frac{1}{1-z}\right)\right]
\end{multline}
we can write $D_l^{(\frac{n-1}{2})}(s_x)$ in terms as
\begin{equation}
D_l^{(\frac{n-1}{2})}(s_x)=R_1(s_x)+R_2(s_x)\cdot\ln\frac{s_x+1}{s_x-1}
\end{equation}
where $R_1(s_x)$ and $R_2(s_x)$ are rational functions of $s_x$.

It is straightforward to check that in the short-distance limit $s_x\rightarrow1$, for both $n$ odd and even, we have
\begin{equation}
D_l^{(\frac{n-1}{2})}(s_x)\rightarrow\frac{\Gamma(\frac{n}{2}-1)}{(s_x-1)^{\frac{n}{2}-1}}\label{short-distance limit}
\end{equation}
which can be recognized as the massless propagator in flat $n$-dimensional space. Thus, eq.\eqref{short-distance limit} shows that we indeed get $\delta(x)$ by hitting $D_l^{(\frac{n-1}{2})}$ with $\square_x$.

One can also check, for the sake of consistency, that $D_l^{(\frac{n-1}{2})}(s_x)$ is the same propagator in AdS$_n$ given in the literature \cite{AdS} (upto a normalization constant $2(2\pi)^\frac{n}{2}$)
\begin{equation}
G_\lambda(\xi)=\frac{\Gamma(\lambda)}{2^{\lambda+1}\pi^{\frac{n-1}{2}}\Gamma(\lambda-\frac{n-3}{2})}
\xi^\lambda F\left(\frac{\lambda}{2},\frac{\lambda+1}{2},\lambda-\frac{n-3}{2},\xi^2\right)
\end{equation}
where $\xi=s_x^{-1}$ and $\lambda=l+n-1$ is the conformal weight of the scalar field. For example, when $n=5$, $l=0$, we have that
\begin{equation}
D_0^{(2)}(s_x)=\sqrt{2\pi}\left[\frac{-2s_x^3+3s_x}{(s_x^2-1)^{\frac{3}{2}}}+2\right]
\end{equation}
which shows the lowest KK mode agrees with the ``dilaton + $i\cdot$ axion'' propagator.

Lastly, let us expand the propagator $\Delta_{2n}$ over $C_l^{(\frac{n-1}{2})}$ and $D_l^{(\frac{n-1}{2})}$. We have shown that
\begin{equation}
-\frac{1}{2}\square\Delta_{2n}=(2\pi)^n\delta(x)\delta(y)
\end{equation}
and
\begin{equation}
-\frac{1}{2}(\square_y+m_l^2)C_l^{(\frac{n-1}{2})}=0,\quad
-\frac{1}{2}(\square_x-m_l^2)D_l^{(\frac{n-1}{2})}=(2\pi)^{\frac{n}{2}}\delta(x)
\end{equation}
where $m_l^2=l(l+n-1)$. Writing that
\begin{equation}
\Delta_{2n}(s_x,s_y)=\sum_{l=0}^\infty N_l D_l^{(\frac{n-1}{2})}(s_x) C_l^{(\frac{n-1}{2})}(s_y)\label{expansion of delta_2n}
\end{equation}
we then have
\begin{equation}
\sum_{l=0}^\infty \frac{2^{-\frac{n}{2}+1}}{\Gamma(\frac{n}{2})} N_l C_l^{(\frac{n-1}{2})}(s_y) (1-s_y^2)^{\frac{n}{2}-1}
=\delta(s_y-1)\label{expansion coefficients (1)}
\end{equation}
On the other hand, from the completeness of $C_l^{(\frac{n-1}{2})}$ we can obtain
\begin{equation}
\sum_{l=0}^\infty \frac{2^{n-2}(l+\frac{n-1}{2})[\Gamma(\frac{n-1}{2})]^2}{\pi\Gamma(n-1)}
 C_l^{(\frac{n-1}{2})}(s_y) (1-s_y^2)^{\frac{n}{2}-1}
=\delta(s_y-1)\label{expansion coefficients (2)}
\end{equation}
Comparing eqs.\eqref{expansion coefficients (1)} and \eqref{expansion coefficients (2)}, we get that
\begin{equation}
N_l=\frac{2^{\frac{n}{2}-1}}{\sqrt\pi}\Gamma\left(\frac{n-1}{2}\right)\left(l+\frac{n-1}{2}\right)
\end{equation}
Thus, eq.\eqref{expansion of delta_2n} can be written as
\begin{equation}
\Delta_{2n}(s_x,s_y)=\sum_{l=0}^\infty
\frac{2^{\frac{n}{2}-1}}{\sqrt\pi}\Gamma\left(\frac{n-1}{2}\right)\left(l+\frac{n-1}{2}\right)
D_l^{(\frac{n-1}{2})}(s_x) C_l^{(\frac{n-1}{2})}(s_y)
\end{equation}
which is our main result of this section.

For more general cases, AdS$_m\times$S$^n(m\neq n)$, AdS$_m\times$S$^m\times\mathbb{R}^n$, etc., the metrics are not conformally flat, so we cannot write the massless scalar propagators compactly like \eqref{propagator in AdS_n x S^n}.

%%%%%%%%%%%%%%%%%%%%%%%%%%%%%%%%%%%%%%%%%%%%%
\section{Coset superspace and covariant derivatives}
In supersymmetry, the configuration space AdS$_5\times$S$^5$ can be regarded as a coset superspace
\begin{equation}
\frac{PSU(2,2|4)}{SO(4,1)\otimes SO(5)}\label{coset (1)}
\end{equation}
By Wick rotations and Lie-algebra identifications, we can bring \eqref{coset (1)} to
\begin{equation}
\frac{GL(4|4)}{[Sp(4)\otimes GL(1)]^2}\label{coset (2)}
\end{equation}
where the superconformal group is represented by unconstrained matrices $GL(4|4)$. Now the spacetime supercoordinates themselves are a representation and the exponential pa-rametrization techniques are no longer necessary.

We write the coset elements as
\begin{equation}
z_A{}^M=\left(\begin{array}{cc}
z_a{}^m & z_a{}^{\bar{m}}\\
z_{\bar{a}}{}^m & z_{\bar{a}}{}^{\bar{m}}
\end{array}\right)
\end{equation}
where the indices $M$ are acted upon by the global superconformal group while the indices $A$ are acted upon by the two isotropy groups $[Sp(4)\otimes GL(1)]^2$. Similarly, the inverse of $z_A{}^M$ can be written as
\begin{equation}
z^{-1}=z_M{}^A=\left(\begin{array}{cc}
z_m{}^a & z_m{}^{\bar{a}}\\
z_{\bar{m}}{}^a & z_{\bar{m}}{}^{\bar{a}}
\end{array}\right)
\end{equation}
which satisfies
\begin{equation}
z_A{}^Mz_M{}^B=\delta_A{}^B, \quad z_M{}^Az_A{}^N=\delta_M{}^N
\end{equation}

In such a matrix representation, the covariant differentiations are simply matrix multiplications. Denoting that
\begin{equation}
(-)^{AB}=\begin{cases}
-1 & \textrm{both }A\textrm{ and }B\textrm{ are fermionic}\\
+1 & \textrm{otherwise}
\end{cases}
\end{equation}
we can define the covariant derivatives of $z_C{}^M$ as follows
\begin{equation}
d_A{}^B z_C{}^M=\delta_C{}^B z_A{}^M (-)^B
\end{equation}
where $(-)^B=(-)^{BB}$. From the fact taht
\begin{equation}
d(z^{-1})=-z^{-1}(dz)z^{-1}
\end{equation}
we can get the covariant derivatives of $z_M{}^C$ as
\begin{equation}
d_A{}^B z_M{}^C=\delta_A{}^C z_M{}^B (-)^{M(A+B)+AB+1}
\end{equation}

For a given field $\Phi=\Phi(z)$ defined on this coset superspace, we say it is chiral if
\begin{equation}
d_a{}^{\bar{b}}\Phi(z)=0
\end{equation}
and antichiral if
\begin{equation}
d_{\bar{a}}{}^b\bar{\Phi}(z)=0
\end{equation}

%%%%%%%%%%%%%%%%%%%%%%%%%%%%%%%%%%%%%%%%%%%%%
\section{Invariant spacetime separations and the propagator}
Now let us consider the scalar propagator of this coset superspace. Here we consider the propagator that is chiral at 1 and antichiral at 2 (while the propagators which are chiral/antichiral on both ends can be determinated by a ``reality condition''), i.e.,
\begin{equation}
\Delta(1,2)=\langle\Phi(1)\bar{\Phi}(2)\rangle
\end{equation}

From the fact that
\begin{equation}
d_a{}^{\bar{b}}(1)z_c{}^M(1)=0=d_a{}^{\bar{b}}(1)z_M{}^{\bar{c}}(1)
\end{equation}
and
\begin{equation}
d_{\bar{a}}{}^b(2)z_M{}^c(2)=0=d_{\bar{a}}{}^b(2)z_{\bar{c}}{}^M(2)
\end{equation}
we can construct the following two globally invariant matrices which have the desired chirality properties
\begin{equation}
M_a{}^b(1,2)=z_a{}^M(1)z_M{}^b(2),\quad
N_{\bar{a}}{}^{\bar{b}}(1,2)=z_{\bar{a}}^M(2)z_M{}^{\bar{b}}(1)\label{M & N}
\end{equation}

For $M_a{}^b(1,2)$, we can contract the local indices with $\Omega_{ab}$, the metric of $Sp(4)$, and mod out the $GL(1)$ charge via $\sqrt{detM}$; then we get the following invariant separation
\begin{equation}
s_x(1,2)=-\frac{1}{4}\frac{tr(\Omega M\Omega M^T)}{\sqrt{detM}}
\equiv\frac{T_x}{\sqrt{D_x}}\label{super s_x (1)}
\end{equation}
where
\begin{equation}
T_x=-\frac{1}{4}tr(\Omega M\Omega M^T)
=\frac{1}{4}\Omega^{ba}\Omega_{dc}M_a{}^c M_b{}^d,\quad D_x=detM
\end{equation}
It is easy to see that
\begin{equation}
d_{\bar{a}}{}^{\bar{b}}s_x=0
\end{equation}
From coset properties, we also have that
\begin{equation}
d_a{}^as_x=0,\quad d_{(ab)}s_x=0
\end{equation}

Similarly, from $N_{\bar{a}}{}^{\bar{b}}(1,2)$ we can construct another invariant separation as
\begin{equation}
s_y(1,2)=-\frac{1}{4}\frac{tr(\Omega N\Omega N^T)}{\sqrt{detN}}
\equiv\frac{T_y}{\sqrt{D_y}}\label{super s_y (1)}
\end{equation}
where
\begin{equation}
T_y=-\frac{1}{4}tr(\Omega N\Omega N^T)
=\frac{1}{4}\Omega^{\bar{b}\bar{a}}\Omega_{\bar{d}\bar{c}}N_{\bar{a}}{}^{\bar{c}}N_{\bar{b}}{}^{\bar{d}},
\quad D_y=detN
\end{equation}
And we also have
\begin{equation}
d_a{}^bs_y=0,\quad d_{\bar{a}}{}^{\bar{a}}s_y=0,\quad d_{(\bar{a}\bar{b})}s_y=0
\end{equation}

Thus, the propagator $\Delta(1,2)$ can be written as a function of $s_x$ and $s_y$
\begin{equation}
\Delta=\Delta(s_x,s_y)
\end{equation}

%%%%%%%%%%%%%%%%%%%%%%%%%%%%%%%%%%%%%%%%%%%%%
\section{KG Equation}
Working under the conventions that
\begin{equation}
\Omega^{ab}\Omega_{ac}=\delta_c{}^b,\quad
\Omega^{\bar{a}\bar{b}}\Omega_{\bar{a}\bar{c}}=\delta_{\bar{c}}{}^{\bar{b}}
\end{equation}
we define the antisymmetric $\Omega$-traceless part of $d_{ab}$ and $d_{\bar{a}\bar{b}}$ as
\begin{equation}
d_{\langle ab\rangle}\equiv\frac{1}{2}(d_a{}^c\Omega_{cb}-d_b{}^c\Omega_{ca})-\frac{1}{4}\Omega_{ab}d_c{}^c
\end{equation}
and
\begin{equation}
d_{\langle\bar{a}\bar{b}\rangle}\equiv
\frac{1}{2}(d_{\bar{a}}{}^{\bar{c}}\Omega_{\bar{c}\bar{b}}-d_{\bar{b}}{}^{\bar{c}}\Omega_{\bar{c}\bar{a}})
-\frac{1}{4}\Omega_{\bar{a}\bar{b}}d_{\bar{c}}{}^{\bar{c}}
\end{equation}
Then the massless KG equation for $\Delta(s_x,s_y)$ is
\begin{equation}
\square(1)\Delta(s_x,s_y)=\left(d_{\langle ab\rangle}^2-d_{\langle\bar{a}\bar{b}\rangle}^2\right)(1)\Delta(s_x,s_y)=0
\label{super KG equation (1)}
\end{equation}

Since
\begin{subequations}\label{KG_x}
\begin{align}
d_{\langle ab\rangle}^2\Delta
=&d^{\langle ab\rangle}d_{\langle ab\rangle}\Delta\\
=&\left(d^{ab}d_{[ab]}-\frac{1}{4}d_a{}^ad_b{}^b\right)\Delta\\
=&d_a{}^bd_b{}^a\Delta
\end{align}
\end{subequations}
and similarly,
\begin{subequations}\label{KG_y}
\begin{align}
d_{\langle\bar{a}\bar{b}\rangle}^2\Delta
=&d^{\langle\bar{a}\bar{b}\rangle}d_{\langle\bar{a}\bar{b}\rangle}\Delta\\
=&\left(d^{\bar{a}\bar{b}}d_{[\bar{a}\bar{b}]}-\frac{1}{4}d_{\bar{a}}{}^{\bar{a}}d_{\bar{b}}{}^{\bar{b}}\right)\Delta\\
=&d_{\bar{a}}{}^{\bar{b}}d_{\bar{b}}{}^{\bar{a}}\Delta
\end{align}
\end{subequations}
we can explicitly write eq.\eqref{super KG equation (1)} as
\begin{equation}\label{super KG equation (2)}
\begin{split}
\square(1)\Delta(s_x,s_y)
=&\left(d_a{}^bd_b{}^a-d_{\bar{a}}{}^{\bar{b}}d_{\bar{b}}{}^{\bar{a}}\right)(1)\Delta(s_x,s_y)\\
=&\left[(d_a{}^bs_x)(d_b{}^as_x)\frac{\partial^2}{\partial s_x^2}+(d_a{}^bd_b{}^as_x)\frac{\partial}{\partial s_x}\right.\\
&\quad\quad\quad\left.-(d_{\bar{a}}{}^{\bar{b}}s_y)(d_{\bar{b}}{}^{\bar{a}}s_y)\frac{\partial^2}{\partial s_y^2}
-(d_{\bar{a}}{}^{\bar{b}}d_{\bar{b}}{}^{\bar{a}}s_y)\frac{\partial}{\partial s_y}\right](1)\Delta(s_x,s_y)
\end{split}
\end{equation}

Now we shall calculate the following quantities in sequence:
\begin{equation*}
d_a{}^b(d_b{}^as_x),\quad
(d_a{}^bs_x)(d_b{}^as_x),\quad
d_{\bar{a}}{}^{\bar{b}}(d_{\bar{b}}{}^{\bar{a}}s_y),\quad
(d_{\bar{a}}{}^{\bar{b}}s_y)(d_{\bar{b}}{}^{\bar{a}}s_y)
\end{equation*}

From the fact that
\begin{equation}
d[\ln(detM)]=tr(M^{-1}dM)
\end{equation}
we can get
\begin{equation}
d_a{}^b\ln D_x=\delta_a{}^b
\end{equation}
thus
\begin{equation}
d_a{}^bs_x=D_x^{-\frac{1}{2}}\left(d_a{}^bT_x-\frac{1}{2}\delta_a{}^bT_x\right)
\end{equation}
Then it is straightforward to obtain
\begin{equation}
d_a{}^b(d_b{}^as_x)=5s_x
\end{equation}

In order to calculate $(d_a{}^bs_x)(d_b{}^as_x)$, we first define the Pfaffians of $\Omega^{ab}$ and $\Omega_{ab}$ as follows
\begin{equation}
Pf(\Omega^{ab})=\frac{1}{8}\epsilon_{abcd}\Omega^{ab}\Omega^{cd}\equiv\alpha
\end{equation}
\begin{equation}
Pf(\Omega_{ab})=\frac{1}{8}\epsilon^{abcd}\Omega_{ab}\Omega_{cd}\equiv\beta
\end{equation}
where $\epsilon$ is the Levi-Civita symbol. From the identity
\begin{equation}
(\alpha\beta)^2=det(\Omega^{ab}\Omega_{ac})=1
\end{equation}
we can consistently choose that
\begin{equation}
\alpha\beta=1
\end{equation}

It is easy to see
\begin{equation}
\alpha\epsilon^{abcd}
=\Omega^{ab}\Omega^{cd}+\Omega^{ca}\Omega^{bd}+\Omega^{ad}\Omega^{bc}
\label{epsilon-Omega}
\end{equation}
Multiplying eq.\eqref{epsilon-Omega} with
\begin{equation*}
\frac{1}{8}(\Omega_{hg}M_a{}^gM_b{}^h)(\Omega_{ji}M_c{}^iM_b{}^j)
\end{equation*}
we find that
\begin{equation}
(d_a{}^bT_x)(d_b{}^aT_x)=2T_x^2-\alpha\beta D_x
\end{equation}
which gives us
\begin{equation}
(d_a{}^bs_x)(d_b{}^as_x)
=\frac{1}{D_x}\left[(d_a{}^bT_x)(d_b{}^aT_x)-(d_a{}^aT_x)T_x+T_x^2\right]
=s_x^2-\alpha\beta
\end{equation}

Similarly, we can obtain the following results for the $s_y$-related quantities
\begin{equation}
d_{\bar{a}}{}^{\bar{b}}(d_{\bar{b}}{}^{\bar{a}}s_y)=5s_y
\end{equation}
and
\begin{equation}
(d_{\bar{a}}{}^{\bar{b}}s_y)(d_{\bar{b}}{}^{\bar{a}}s_y)=s_y^2-\alpha^\prime\beta^\prime
\end{equation}
where $\alpha^\prime$ is the Pfaffian of $\Omega^{\bar{a}\bar{b}}$ and $\beta^\prime$ is the Pfaffian of $\Omega_{\bar{a}\bar{b}}$
\begin{equation}
Pf(\Omega^{\bar{a}\bar{b}})
=\frac{1}{8}\epsilon_{\bar{a}\bar{b}\bar{c}\bar{d}}\Omega^{\bar{a}\bar{b}}\Omega^{\bar{c}\bar{d}}
\equiv\alpha^\prime
\end{equation}
\begin{equation}
Pf(\Omega_{\bar{a}\bar{b}})
=\frac{1}{8}\epsilon^{\bar{a}\bar{b}\bar{c}\bar{d}}\Omega_{\bar{a}\bar{b}}\Omega_{\bar{c}\bar{d}}
\equiv\beta^\prime
\end{equation}
and we can choose
\begin{equation}
\alpha^\prime\beta^\prime=1
\end{equation}

Substituting the above results into the KG equation \eqref{super KG equation (2)}, we get that
\begin{equation}
\square\Delta(s_x,s_y)
=\left[(s_x^2-1)\frac{\partial^2}{\partial s_x^2}+5s_x\frac{\partial}{\partial s_x}
-(s_y^2-1)\frac{\partial^2}{\partial s_y^2}-5s_y\frac{\partial}{\partial s_y}\right]\Delta(s_x,s_y)=0
\label{super KG equation (3)}
\end{equation}
which is of the same form as eq.\eqref{KG equation in AdS_n x S^n (2)}. Thus, the propagator we have here is of the same form as before (though now $s_x$ and $s_y$ are different)
\begin{equation}
\Delta(s_x,s_y)=\frac{3!}{(s_x-s_y)^4}\label{propagator in the coset space}
\end{equation}

%%%%%%%%%%%%%%%%%%%%%%%%%%%%%%%%%%%%%%%%%%%%%
\section{$\kappa$-symmetry constraint}
One might wonder whether the propagator \eqref{propagator in the coset space} also satisfies the $\kappa$-symmetry constraint as it should
\begin{equation}
\kappa_{\bar{b}a}(1)\Delta(s_x,s_y)
\equiv\left[d_a{}^c d_{\bar{b}c}+\eta d_{\bar{b}}{}^{\bar{c}}d_{\bar{c}a}+\chi d_{\bar{b}a}\right](1)
\Delta(s_x,s_y)=0\label{kappa-symmetry constraint (1)}
\end{equation}
where $\eta=\pm1$ and $\chi=$ constant. The $\chi d_{\bar{b}a}$ term is due to the ordering ambiguity of $d_a{}^c$ and $d_{\bar{b}c}$ as well as $d_{\bar{b}}{}^{\bar{c}}$ and $d_{\bar{c}a}$.

Firstly, we would like to show that the KG equation \eqref{super KG equation (2)} can be derived from the $\kappa$-symmetry constraint \eqref{kappa-symmetry constraint (1)}.

The chirality property of $\Delta(s_x,s_y)$ at point 1 implies
\begin{equation}
d^{a\bar{b}}(1)\Delta(s_x,s_y)=0
\end{equation}
Together with \eqref{kappa-symmetry constraint (1)}, we have that
\begin{equation}
\{d^{a\bar{b}},\kappa_{\bar{b}a}\}(1)\Delta(s_x,s_y)=0
\end{equation}

It is straightforward to prove the following (anti-)commutation relation
\begin{equation}
[d_A{}^B,d_C{}^D\}
=(-)^B\delta_C{}^B d_A{}^D-(-)^{B(C+D)+CD}\delta_A{}^D d_C{}^B
\end{equation}

After some algebra, we can get
\begin{equation}
\{d^{a\bar{b}},\kappa_{\bar{b}a}\}
=4d^{ab}d_{ab}-4\eta d^{\bar{a}\bar{b}}d_{\bar{a}\bar{b}}
+(\eta-1)d_a{}^ad_{\bar{b}}{}^{\bar{b}}+4(1+\chi-4\eta)(d_a{}^a-d_{\bar{b}}{}^{\bar{b}})
\end{equation}
thus
\begin{equation}
\{d^{a\bar{b}},\kappa_{\bar{b}a}\}\Delta
=4\left(d^{ab}d_{[ab]}-\eta d^{\bar{a}\bar{b}}d_{[\bar{a}\bar{b}]}\right)\Delta
\end{equation}
Comparing with eqs.\eqref{KG_x} and \eqref{KG_y}, we recognize that the right-hand side is the KG equation piece $4\square\Delta$. (As a consistency check, $\eta$ is left unfixed here and will be fixed at the end.)

Now we shall explicitly calculate $\kappa_{\bar{b}a}(1)\Delta(s_x,s_y)$ and see that there is no more information contained in the $\kappa$-symmetry constraint besides the KG equation.

First of all, from eq.\eqref{kappa-symmetry constraint (1)} it is easy to get that
\begin{equation}
\begin{split}
\kappa_{\bar{b}a}(1)\Delta(s_x,s_y)
=&\left\{(d_a{}^cs_x)(d_{\bar{b}c}s_x)\frac{\partial^2}{\partial s_x^2}
+[d_a{}^c(d_{\bar{b}c}s_x)+(\chi-4\eta)d_{\bar{b}a}s_x]\frac{\partial}{\partial s_x}\right.\\
&+\eta(d_{\bar{b}}{}^{\bar{c}}s_y)(d_{\bar{c}a}s_y)\frac{\partial^2}{\partial s_y^2}
+[\eta d_{\bar{b}}{}^{\bar{c}}(d_{\bar{c}a}s_y)+(\chi+1)d_{\bar{b}a}s_y]\frac{\partial}{\partial s_y}\\
&+\left.[(d_a{}^cs_x)(d_{\bar{b}c}s_y)+\eta(d_{\bar{b}}{}^{\bar{c}}s_y)(d_{\bar{c}a}s_x)]
\frac{\partial^2}{\partial s_x \partial s_y}\right\}(1)\Delta(s_x,s_y)\label{kappa-symmetry constraint (2)}
\end{split}
\end{equation}

Using the same methods as before, we get the following results for the first line of eq.\eqref{kappa-symmetry constraint (2)}
\begin{equation}
(d_a{}^cs_x)(d_{\bar{b}c}s_x)=\frac{1}{4}(d_{\bar{b}a}\ln D_x)(s_x^2-1)
\end{equation}
\begin{equation}
d_a{}^c(d_{\bar{b}c}s_x)=3D_x^{-\frac{1}{2}}d_{\bar{b}a}T_x-\frac{1}{4}(d_{\bar{b}a}\ln D_x)s_x
\end{equation}
and for the second line of eq.\eqref{kappa-symmetry constraint (2)}
\begin{equation}
(d_{\bar{b}}{}^{\bar{c}}s_y)(d_{\bar{c}a}s_y)=\frac{1}{4}(d_{\bar{b}a}\ln D_y)(s_y^2-1)
\end{equation}
\begin{equation}
d_{\bar{b}}{}^{\bar{c}}(d_{\bar{c}a}s_y)=-2D_y^{-\frac{1}{2}}d_{\bar{b}a}T_y+\frac{9}{4}(d_{\bar{b}a}\ln D_y)s_y
\end{equation}

The following identity is useful
\begin{equation}
\begin{split}
0=&z_{\bar{a}}{}^M(2)z_M{}^b(2)\\
=&z_{\bar{a}}{}^M(2)[z_M{}^cz_c{}^N+z_M{}^{\bar{c}}z_{\bar{c}}{}^N](1)z_N{}^b(2)\\
=&[z_{\bar{a}}{}^M(2)z_M{}^c(1)]M_c{}^b+N_{\bar{a}}{}^{\bar{c}}[z_{\bar{c}}{}^N(1)z_N{}^b(2)]
\end{split}
\end{equation}
from which we can prove that
\begin{equation}
(d_{\bar{b}}{}^{\bar{c}}T_y)(d_{\bar{c}a}T_x)=-(d_a{}^cT_x)(d_{\bar{b}c}T_y)
\end{equation}
and
\begin{equation}
d_{\bar{b}a}\ln D_y=-d_{\bar{b}a}\ln D_x
\end{equation}
Thus, the third line of eq.\eqref{kappa-symmetry constraint (2)} can be simplified as
\begin{equation}
\begin{split}
(d_a{}^cs_x)(d_{\bar{b}c}s_y)+\eta(d_{\bar{b}}{}^{\bar{c}}s_y)(d_{\bar{c}a}s_x)
=&(\eta-1)\left[-(D_xD_y)^{-\frac{1}{2}}(d_a{}^cT_x)(d_{\bar{b}c}T_y)\right.\\
&+\frac{1}{2}D_y^{-\frac{1}{2}}(d_{\bar{b}a}T_y)s_x-\frac{1}{2}D_x^{-\frac{1}{2}}(d_{\bar{b}a}T_x)s_y\\
&+\left.\frac{1}{4}(d_{\bar{b}a}\ln D_x)s_xs_y\right]
\end{split}
\end{equation}

Substituting all these results into $\kappa$-symmetry constraint \eqref{kappa-symmetry constraint (2)}, we get that
\begin{equation}
\begin{split}
\kappa_{\bar{b}a}(1)\Delta(s_x,s_y)
=&\left\{(\chi-4\eta+3)D_x^{-\frac{1}{2}}d_{\bar{b}a}T_x\frac{\partial}{\partial s_x}
+(\chi-2\eta+1)D_y^{-\frac{1}{2}}d_{\bar{b}a}T_y\frac{\partial}{\partial s_y}\right.\\
&+(\eta-1)\left[-(D_xD_y)^{-\frac{1}{2}}(d_a{}^cT_x)(d_{\bar{b}c}T_y)
+\frac{1}{2}D_y^{-\frac{1}{2}}(d_{\bar{b}a}T_y)s_x\right.\\
&\quad\quad\left.-\frac{1}{2}D_x^{-\frac{1}{2}}(d_{\bar{b}a}T_x)s_y
+\frac{1}{4}(d_{\bar{b}a}\ln D_x)s_xs_y\right]\frac{\partial^2}{\partial s_x \partial s_y}\\
&+\frac{1}{4}(d_{\bar{b}a}\ln D_x)\left[(s_x^2-1)\frac{\partial^2}{\partial s_x^2}
-(2\chi-8\eta+1)s_x\frac{\partial}{\partial s_x}\right.\\
&\quad\quad\left.\left.-\eta(s_y^2-1)\frac{\partial^2}{\partial s_y^2}
-(-2\chi+9\eta-2)s_y\frac{\partial}{\partial s_y}\right]\right\}(1)\Delta(s_x,s_y)
\label{kappa-KG}
\end{split}
\end{equation}
It is not difficult to see that the last two lines in eq.\eqref{kappa-KG} is the KG equation piece. Comparing with eq.\eqref{super KG equation (3)}, we can find that
\begin{equation}
\eta=1,\quad\chi=1
\end{equation}
Plugging it back into eq.\eqref{kappa-KG}, we see that all the other terms automatically vanish, which shows the result is consistent.

%%%%%%%%%%%%%%%%%%%%%%%%%%%%%%%%%%%%%%%%%%%%%
\section{$s_x$ and $s_y$ in an explicit coset parametrization}
We shall now choose an explicit parametrization of the coset superspace \eqref{coset (2)} and derive the corresponding expressions of $s_x$ defined in eq.\eqref{super s_x (1)} and $s_y$ defined in eq.\eqref{super s_y (1)}.

We write the relevant parts of $z$ and $z^{-1}$ in chiral representation as
\begin{equation}
z_a{}^M=x_a{}^n(\delta_n{}^m,\theta_n{}^{\bar{m}}),\quad
z_M{}^{\bar{a}}=(-\theta_m{}^{\bar{n}},\delta_{\bar{m}}{}^{\bar{n}})y_{\bar{n}}{}^{\bar{a}}
\end{equation}
and in antichiral representation as
\begin{equation}
z_{\bar{a}}{}^M=y_{\bar{a}}{}^{\bar{n}}(-\bar{\theta}_{\bar{n}}{}^m,\delta_{\bar{n}}{}^{\bar{m}}),\quad
z_M{}^a=(\delta_m{}^n,\bar{\theta}_{\bar{m}}{}^n)x_n{}^a
\end{equation}
where $x_n{}^a$ is the inverse of $x_a{}^n$ and $y_{\bar{n}}{}^{\bar{a}}$ is the inverse of $y_{\bar{a}}{}^{\bar{n}}$, i.e.,
\begin{equation}
x_a{}^nx_n{}^b=\delta_a{}^b,\quad y_{\bar{a}}{}^{\bar{n}}y_{\bar{n}}{}^{\bar{b}}=\delta_{\bar{a}}{}^{\bar{b}}
\end{equation}

Recalling the definition of $M_a{}^b(1,2)$ in eq.\eqref{M & N}, we can use chiral representation at point 1 and antichrial representation at point 2 to get
\begin{equation}
M_a{}^b(1,2)=x_a{}^m(1)[I+\theta(1)\bar{\theta}(2)]_m{}^nx_n{}^b(2)
\end{equation}
Then $s_x$ can be expressed in terms of $x$ and $\theta,\bar{\theta}$:
\begin{equation}
s_x(1,2)=-\frac{1}{4}\frac{tr(\Omega M(1,2)\Omega M^T(1,2))}{\sqrt{detM(1,2)}}
\label{super s_x (2)}
\end{equation}

To simplify eq.\eqref{super s_x (2)}, we define the following $4\times4$ antisymmetric matrix
\begin{equation}
X^{mn}=\Omega^{ba}x_a{}^mx_b{}^n
\end{equation}
One can readily get its invese
\begin{equation}
(X^{-1})_{mn}=-\Omega_{ba}x_m{}^ax_n{}^b
\end{equation}

We regard $X^{mn}$ as a 6-dimensional vector and lower its indices as follows
\begin{equation}
X_{mn}\equiv\frac{1}{2}\epsilon_{mnpq}X^{pq}
=-Pf(X)(X^{-1})_{mn}
\end{equation}
Correspondly, we define the inner product
\begin{equation}
X\cdot X^\prime\equiv\frac{1}{4}X^{mn}X_{nm}^\prime
\end{equation}
and get the norm of $X$
\begin{equation}
||X||=\sqrt{X\cdot X}=\sqrt{-Pf(X)}
\end{equation}
Thus, we can write 6-dimensional unit vectors as follows
\begin{equation}
\hat{X}^{mn}=\frac{X^{mn}}{\sqrt{-Pf(X^{mn})}}\label{X_hat}
\end{equation}

We also define
\begin{equation}
\Theta_m{}^n(1,2)\equiv\frac{(I+\theta(1)\bar{\theta}(2))_m{}^n}{[det(I+\theta(1)\bar{\theta}(2))]^{\frac{1}{4}}}\label{Theta}
\end{equation}

Using the notation defined in eqs.\eqref{X_hat} and \eqref{Theta}, eq.\eqref{super s_x (2)} can be written as
\begin{equation}
s_x(1,2)=\frac{1}{4}tr\left[\hat{X}(1)\Theta(1,2)\hat{X}^{-1}(2)\Theta^T(1,2)\right]
\end{equation}
When $\theta,\bar{\theta}=0$, i.e, $\Theta=I$, we get that
\begin{equation}
s_x(1,2)=\frac{1}{4}tr\left[\hat{X}(1)\hat{X}^{-1}(2)\right]=-\hat{X}(1)\cdot\hat{X}^{-1}(2)
\end{equation}

Similarly, for $s_y$ we get that
\begin{equation}
N_{\bar{a}}{}^{\bar{b}}(1,2)
=y_{\bar{a}}{}^{\bar{m}}(2)[I+\bar{\theta}(2)\theta(1)]_{\bar{m}}{}^{\bar{n}}y_{\bar{n}}{}^{\bar{b}}(1)
\end{equation}
and then
\begin{equation}
s_y(1,2)=-\frac{1}{4}\frac{tr(\Omega N(1,2)\Omega N^{T}(1,2))}{\sqrt{det\ N(1,2)}}\label{super s_y (2)}
\end{equation}

We define the following $4\times4$ antisymmetric matrix
\begin{equation}
Y^{\bar{m}\bar{n}}=\Omega^{\bar{b}\bar{a}}y_{\bar{a}}{}^{\bar{m}}y_{\bar{b}}{}^{\bar{n}}
\end{equation}
whose inverse can be written as
\begin{equation}
(Y^{-1})_{\bar{m}\bar{n}}=-\Omega_{\bar{b}\bar{a}}y_{\bar{m}}{}^{\bar{a}}y_{\bar{n}}{}^{\bar{b}}
\end{equation}
We also regard $Y^{\bar{m}\bar{n}}$ as a 6-dimensional vector. We can lower the indices, define the inner product and get the norm of $Y$ similarly as before. Then we define the following 6-dimensional unit vector 
\begin{equation}
\hat{Y}^{\bar{m}\bar{n}}=\frac{Y^{\bar{m}\bar{n}}}{\sqrt{Pf(Y^{\bar{m}\bar{n}})}}\label{eq:Y_hat}
\end{equation}

Here we also define
\begin{equation}
\bar{\Theta}_{\bar{m}}{}^{\bar{n}}(1,2)
\equiv\frac{(I+\bar{\theta}(2)\theta(1))_{\bar{m}}{}^{\bar{n}}}{[det(I+\bar{\theta}(2)\theta(1))]^{\frac{1}{4}}}
\label{eq:Theta_bar}
\end{equation}

Using the notation defined in eqs.\eqref{eq:Y_hat} and \eqref{eq:Theta_bar}, eq.\eqref{super s_y (2)} can be written as
\begin{equation}
s_{y}=\frac{1}{4}tr\left[\hat{Y}(2)\bar{\Theta}(1,2)\hat{Y}^{-1}(1)\bar{\Theta}^{T}(1,2)\right]
\end{equation}
When $\theta,\bar{\theta}=0$, i.e., $\bar{\Theta}=I$, we have
\begin{equation}
s_{y}(1,2)=\frac{1}{4}tr[\hat{Y}(2)\hat{Y}^{-1}(1)]=\hat{Y}(1)\cdot\hat{Y}(2)
\end{equation}

%%%%%%%%%%%%%%%%%%%%%%%%%%%%%%%%%%%%%%%%%%%%%
\section{Conclusions}
We have shown that the (bulk-to-bulk) superspace propagator is much simpler than the component propagators following from expansion in the fermionic coordinates and modes of S${}^5$. By choosing a convenient coordinate system and taking the corresponding limit, one can derive the bulk-to-boundary propagator. 

However, the propagator we gave was for the field strength, not the prepotential, which is necessary for describing interactions.  It may be simplest to use a particular lightcone gauge for this purpose, since the color-singlet operators of the boundary CFT correspond to on-shell states in the bulk; this naturally reduces the number of bosonic and fermionic coordinates of 10D IIB supergravity to those of 4D N=4 Yang-Mills theory.  We are now investigating the propagator for the prepotential in lightcone gauge and will provide details in future publications.

\section*{Note added}
After submission of this work to arxiv, the authors were informed about an earlier work \cite{Dorn} which gives most of the bosonic results for the scalar propagator in AdS$_n\times$S$^n$ (except the expressions of the general AdS$_n$ propagator in terms of rational functions, and logarithms for $n$ even).

%%%%%%%%%%%%%%%%%%%%%%%%%%%%%%%%%%%%%%%%%%%%%
\acknowledgments
WS thanks Leonardo Rastelli for discussions.  Research was supported in part by NSF grant No.\ PHY-0653342.
RNH is grateful for the hospitality of YITP and his visit is supported by the China Scholarship Council (No. 2008102156).

%%%%%%%%%%%%%%%%%%%%%%%%%%%%%%%%%%%%%%%%%%%%%

\end{document}